\documentclass[showpacs,showkeys,amsmath,amssymb,superscript, 
twocolumn]{revtex4-1}

\usepackage{amsopn}
\usepackage{amsmath}
\usepackage{amssymb}
\usepackage{hyperref}
\usepackage{graphicx}
\usepackage{color}
\usepackage{listings}
\usepackage{subfig}
\usepackage{enumerate}
\usepackage[ruled,vlined]{algorithm2e}

\providecommand{\keywords}[1]{\newline Keywords: #1}

\providecommand{\pacs}[1]{\newline PACS: #1}

\newcommand{\Complex}{\ensuremath{\mathbb{C}}}

\newcommand{\boldzero}{\ensuremath{\mathbf{0}}}

\newcommand{\ket}[1]{\ensuremath{|#1\rangle}}
\newcommand{\bra}[1]{\ensuremath{\langle#1|}}

\newcommand{\braket}[2]{\ensuremath{\langle{#1}|{#2}\rangle}}

\newcommand{\1}{{\rm 1\hspace{-0.9mm}l}}
\newcommand{\Id}{{\rm 1\hspace{-0.9mm}l}}

\newtheorem{lemma}{Lemma}

\newtheorem{fact}{Fact}

\begin{document}
\title{Quantum spatial search on planar networks}
\author{Przemys{\l}aw Sadowski}
\email{psadowski@iitis.pl}
\affiliation{Institute of Theoretical and Applied Informatics, Polish Academy
of Sciences, Ba{\l}tycka 5, 44-100 Gliwice, Poland}

\keywords{Quantum walk; Quantum search algorithm; Apollonian networks}
\pacs{03.67.Ac; 03.65.Aa; 03.67.Lx; 02.10.Ox; 89.75.Fb}

\begin{abstract}
This paper deals with the problem of the requirements for quantum systems that
enable one to design efficient quantum algorithms. We rise the issue of the
possibility to utilise the non-complete networks for algorithmic purposes.
In particular we consider applications for the spatial search problems. We
focus on showing that the asymptotic complexity widely discussed in the related
work is not enough tool for determining the potential of the network. We
provide an example of a network where the asymptotic complexity is the same for
a variety of cases and yet it is not always possible to implement successful
search procedure within the quantum walk scheme.

The examples are based on an Apollonian network which models a variety of 
iteratively generated planar networks. The network is planar, exhibits 
linear growth of edges number, consists nodes of different degrees and has the 
small-world and scale-free properties. This motivates its analysis due to the 
simplicity in terms of connections density and potential for quantum phenomena 
due to the nodes diversity. 

\end{abstract}

\maketitle

\section{Introduction}
One of the key motivations for the development of quantum information theory 
is the possibility to utilize quantum systems for computational 
purposes.
Main progress in this field has been done since two most known algorithms were 
introduced -- Grover's search and Shor's factorization algorithms.
In this work we study the issue of the of computational resources 
necessary to perform effective computational tasks in quantum systems.
The issue has been studied in terms of the possible interactions represented by 
a connection graph~\cite{mater2013blueprint, konno2013spectral, 
watrous2001undirected, konno2013spectral},
most commonly within quantum walk 
notation \cite{portugal2013quantum}.

The original Grover's search algorithm corresponds to an underlying complete 
graph of connections~\cite{portugal2013quantum}.
Related results considering the quantum search on less complex networks are 
based on such graphs as hypercubes $2^n$~\cite{shenvi2003walksearch}, twisted 
toroids~\cite{douglas2013complexity}, Hanoi 
networks~\cite{marquezino2013hanoi} and other.
These graphs are not planar.
Moreover, the networks exhibit high symmetry and in particular in each of the 
cases all of the nodes can be considered equivalent.
The results regarding the effectiveness of the quantum walk search on general 
networks are based on asymptotic complexity 
analysis~\cite{ambainis2007quantum}.

Our goal is to find a network suitable for efficient computation with 
properties similar to the 2-dimensional grid which is planar and sparse.
We turn our attention onto iteratively generated planar networks.
For this purpose we study Apollonian networks.
These networks are planar and the average node degree converges to a constant 
number. Additionally, the nodes in the network are not equivalent (even in the 
case of the last generation nodes only), what could allow quantum phenomena 
to occur~\cite{faccin2013degree}.

In this work we focus on showing that the asymptotic search complexity is not 
enough to determine the possibility to successfully implement search algorithm 
and when it is possible to efficiently utilise Apollonian network for search 
purposes.
In particular, we consider two cases of possible marked nodes location regions 
and show that they lead to the same asymptotic results.
However, when we consider the possibility to implement the search with single 
measurement we show that the basic search approach fails but the appropriate 
use of the symmetry in the network 
allows us to maintain the optimal quantum search speed-up.

\subsection{Apollonian networks}\label{sec:apollonian}

We focus on an example of highly symmetric iteratively generated planar network 
-- Apollonian network -- an undirected graph constructed during the process of 
filling the plane with circles~\cite{boyd73osculatory,boyd73residual}. 
Namely, we consider a graph representing connections between the circles. Each 
node represents one circle and is connected with nodes that represent 
neighbouring circles. While adding new circle to the plane a new node is 
created with appropriate connections.
Equivalently it can be defined as the result of 
subdivision of a triangle into smaller triangles, which is the case of three 
initial circles on the plane.
The model is also considered with various modifications.
In particular random Apollonian networks are created by dividing only a number 
of triangles in the network.
A model defined this way can be used to describe 
a vast variety of iteratively generated planar networks.
Any iteratively generated family of planar networks where each new 
node created during creation process is connected with all possible neighbours 
can be considered as Apollonian network.
Apollonian networks have attracted extensive attention in the field of 
studying large complex networks due to their small-world and scale-free 
properties. For more detailed analysis of the properties of Apollonian networks 
we refer to~\cite{doye05self-similar}.

\begin{figure}
\includegraphics[scale=0.5]{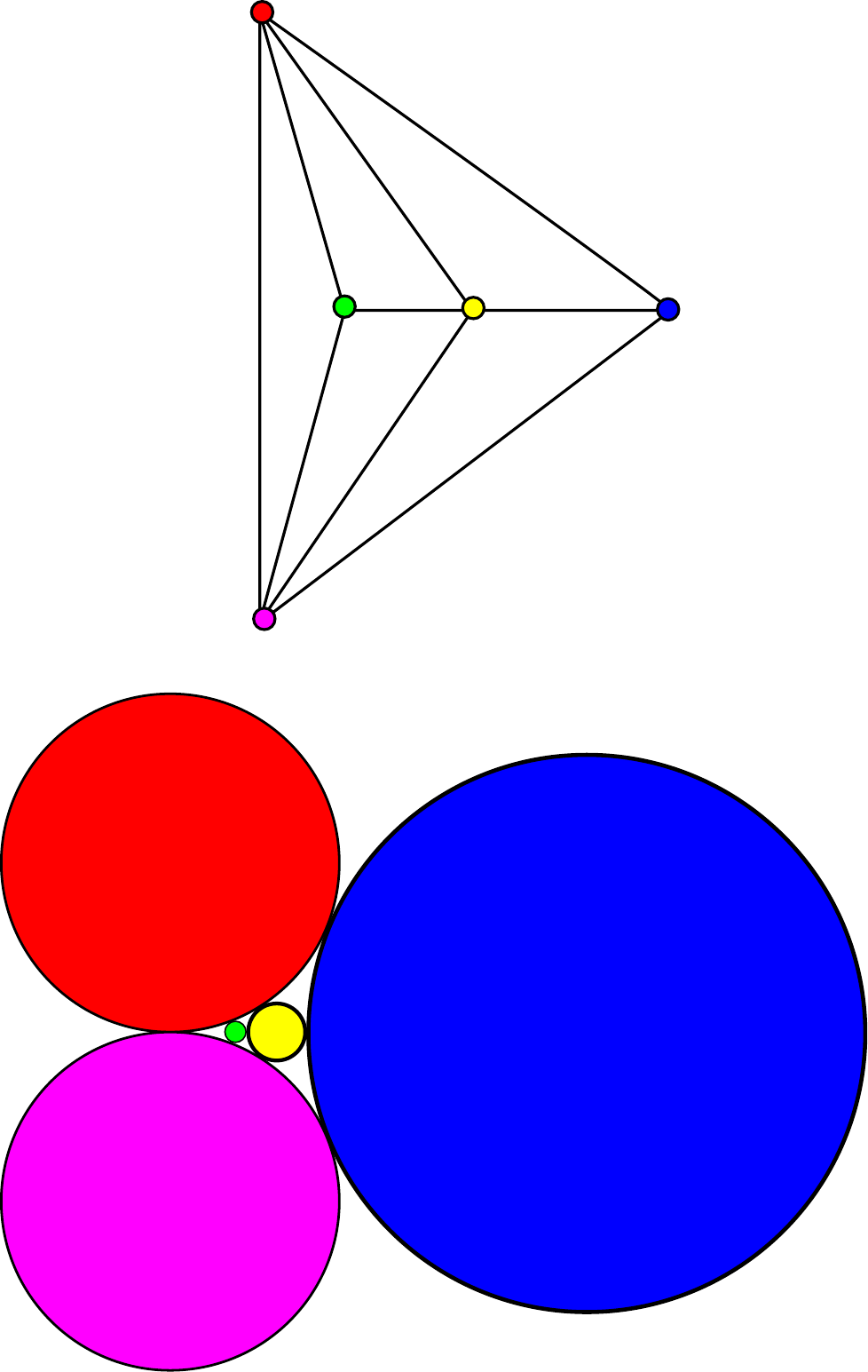}
\caption{Correspondence between filling a plane with circles and generating 
an Apollonian network.}\label{fig:apollonian_svg}
\end{figure}

\begin{figure*}[!h]
\centering
\includegraphics[width=0.6\textwidth]{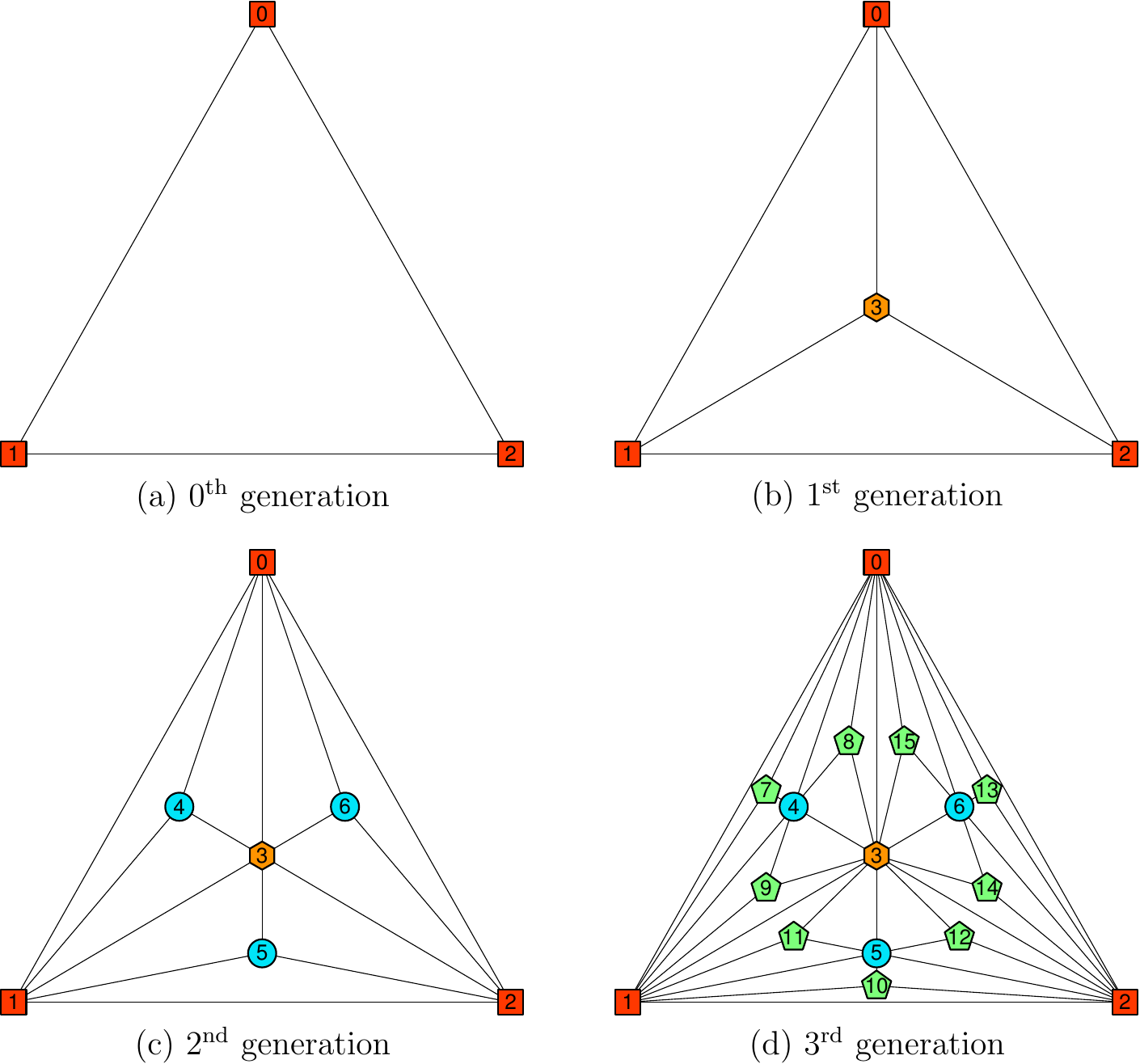}
\caption{ {\bf An illustration of the construction of an Apollonian network.}
Red squares illustrate the nodes in the 0\textsuperscript{th} generation, an
orange hexagon in the 1\textsuperscript{st} generation, blue circles in the
2\textsuperscript{nd} generation and green pentagons in the
3\textsuperscript{rd} generation.}
\label{fig:example-network}
\end{figure*}

The main motivation to execute spatial search algorithm on Apollonian network 
is the network complexity. Basic version of the Grover's algorithm 
requires a complete network to run \cite{portugal2013quantum}. This means that 
the system with $N$ positions must support $\binom{N}{2}=N(N-1)/2=O(N^2)$ 
interactions. Additionally, the degree of all nodes is equal to $N-1$.

In the case of Apollonian network the total number of connections increases 
linearly and is equal to 
\begin{equation}
Edges(N) = 2+\sum_{k=0}^{K} 3^k = \frac{3}{2}(3^{K}+1) = O(N),
\end{equation}
where the number of nodes in the network $N$ is related to the 
generation $K$ of the network by the formula
\begin{equation}
N=\frac{1}{2}(3^K+5).\label{eq:number-of-nodes}
\end{equation}
Additionally, maximum node degree in the network is
$$d_{max}(N) = 3\cdot2^{K-1}= O(N^{\log_3 2})\approx O(N^{0.63})$$
while the average node degree is equal to $ (6-\frac{24}{3^K+5}) $ and 
converges to 6.

\section{Quantum walk search on Apollonian network}\label{sec:search}

In this section we aim at introducing a quantum walk model.
We consider asymptotic search complexity and discuss possibility to implement 
the search with a single measurement. We consider two cases and provide a 
comparison.

\def\connected{\tilde{-}}
\subsection{The model for the quantum walk algorithm}\label{sec:model}
Quantum walks model has been introduced as a modification of classical random 
walks \cite{zagury1993walks}. Most attention was paid to quantum walks on 
a line, a lattice and other regular graphs~\cite{kempe2003overview, 
aharonov2001ongraphs, childs2004spatialsearch}.

In the usual quantum walk model we consider a bipartite system 
$\mathcal{H_C}\otimes\mathcal{H_P}$.
Every step of the evolution $U$ is a composition of coin and 
shift operators
\begin{equation}
\ket{\psi_{t+1}} = U \ket{\psi_t} = S(C\otimes\1) \ket{\psi_t},
\end{equation}
where $\ket{\psi_0}\in\mathcal{H_C}\otimes\mathcal{H_P}$ is the initial state.
In this case the coin operator acts on the internal state subspace. Because we 
consider a regular structure the internal state space is the same for each 
position. Thus, the overall operator can be represented as a tensor product.

When the considered network is not regular we need to modify the usual product
space model. We consider all the possible pairs $(j,i)$, where node $i$ is
connected to node $j$. Let us denote the degrees of all of the nodes in the
network by $d_1, ..., d_N$ such that the degree of $i$-th node is equal to
$d_i$. We construct a space of dimension equal to $\sum d_i$ with basis vectors
$\{\ket{j,i}\}_{i\connected j}$, where $\connected$ denotes the relation
between joined nodes. The space can be represented as a direct sum of subspaces
corresponding to the subsequent nodes. We attribute every position $i$ with a
space $D_i=\Complex^{d_i}$ and consider a quantum system defined as
\def\statespace{\mathcal{X}}
\begin{equation}
\statespace=D_0\oplus\cdots\oplus D_{{N-1}}.
\label{eq:directsumspace}
\end{equation}
We denote basis states by $\ket{j, 
i}=\boldzero_0\oplus\ldots\oplus\boldzero_{i-1}\oplus\ket{j}\oplus
\boldzero_{i+1}\oplus\ldots\oplus\boldzero_{N-1}$
which represent a state with position equal to $i$ directed to node $j$.
The shift operator in this walk is defined as
\begin{equation}
S = \sum_{i}\sum_{j\tilde{-}i} \ket{j, i}\bra{i, j}.
\end{equation}
Additionally, before each shift the coin operator is applied.
In the general case, when $C_i$ denotes the coin operator for $i$-th node, the 
overall operator denotes
\begin{equation}
C = C_0\oplus C_1\oplus\ldots\oplus C_{N-1}.
\end{equation}
In order to execute the walk performing the search procedure we set $C_i = 
G_{d_i}$, where $G_d$ is a 
$d$-dimensional Grover's diffusion operator 
\begin{equation}
G_N = \left\{\frac{2}{N}\right\}_{i,j}-\Id.
\label{eq:grover_operator}
\end{equation}
The only exception is the marked node where the coin operator is multiplied by 
$(-1)$ such that $C_m = -G_{d_m}$.

In this work for the search purposes we assume that the initial state 
$\ket{\psi_0}$ is an equal 
superposition over a fixed set of basis states. In particular we consider the 
whole basis and the states with position representing nodes form the last 
generation of a given network.

\subsection{Asymptotic complexity}\label{sec:walk-asymptotic}

In this section we recall the results that imply that searching for an 
arbitrary node in a network within the introduced quantum walk model is 
possible with complexity $O(\sqrt{N})$. We stress that this results do not give 
any formula for the exact time step to perform the measurement and consider 
each of the marked states separately thus do not determine whether it is 
possible to choose a common measurement step time that succeeds for marking any 
of the nodes.

\begin{lemma}[Lemma 3 in \cite{ambainis2007quantum}]
Let $\mathcal{H}$ be a finite dimensional Hilbert space and 
$\ket{\psi_1},\ldots,\ket{\psi_m}$  be an orthonormal basis for $\mathcal{H}$.
Let $\ket{\psi_{good}},\ket{\psi_{start}}$ be two states in $\mathcal{H}$ 
which are superpositions of $\ket{\psi_1},...,\ket{\psi_m}$ with real 
amplitudes and
$\braket{\psi_{good}}{\psi_{start}}=\alpha$. Let $U_1, U_2$ be unitary 
transformations on $\mathcal{H}$ with the following properties:
\begin{enumerate}
\item $U_1$ is the transformation that flips the phase on $\ket{\psi_{good}}$ 
($U_1 \ket{\psi_{good}} = -\ket{\psi_{good}}$ and leaves any state
orthogonal to $\ket{\psi_{good}}$ unchanged.

\item $U_2$ is a transformation which is described by a real-valued $m \times 
m$ 
matrix in the basis $\ket{\psi_1},...,\ket{\psi_m}$.
Moreover, $U_2 \ket{\psi_{start}} = \ket{\psi_{start}}$ and, if $\ket{\psi}$ is 
an eigenvector of $U_2$ perpendicular to $\ket{\psi_{start}}$, then
$U_2\ket{\psi} = e^{i\theta}\ket{\psi}$ for $\theta \in [\sigma, 2\pi - 
\sigma]$ (where $\sigma$ is a constant, $\sigma>0$)
\end{enumerate}
Then, there exists $t = O( 1/\alpha )$ such that
$|\bra{\psi_{good}} (U_2 U_1 )^t \ket{\psi_{start}}|=\Omega(1)$.
(The constant under $\Omega(1)$ is independent of $\alpha$ but can depend on 
$\sigma$.)
\end{lemma}

In this paper we consider the space given in Eq. (\ref{eq:directsumspace}) with 
basis states $\{\ket{j,i}\}$ and operators $U_1=O_m$, $U_2=SC$ which are real 
in the considered basis.
The operator $O_m$ satisfies property 1 trivially.
We prove property 2 for the operator $SC$ on a restricted state space with use 
of the following fact.

\begin{fact}
Lets denote the subspace spanned by the $\lambda$-eigenvectors of $U_2$ for 
$\lambda\ne1$ by $\overline{\statespace_1}$.
The subspace $\statespace'<\statespace$ spanned by the states from 
$\overline{\statespace_1}\cup\{\ket{\psi_{start}}\}$ satisfies:
\begin{enumerate}[a)]
\item $\ket{\mathrm{\psi_{start}}}\in\statespace'$,
\item $\ket{\mathrm{t_m}}\in\statespace'$, $m=1,\ldots,N$,
\item $SCO_m\ket{\mathrm{\psi}}\in\statespace',$ for 
$\ket{\psi}\in\statespace'$, $m=1,\ldots,N$.
\end{enumerate}
\end{fact}

Pt. a) is satisfied trivially.
The marked state $\ket{t_m}$ belongs to 
$\statespace'$ because is is assumed to be a local equal superposition.
The states not included in $\statespace'$ are exactly 1-eigenvectors orthogonal 
to the initial state and can be expressed as states 
that are orthogonal to any local equal superposition.
It follows from the fact that  the $U_2$ operator is a composition of two 
operators that exhibits only $\pm1$ eigenvalues.
Thus,  any 1-eigenvector of 
$U_2$ is a common 1 (or -1) eigenvector of $S$ and $C$. 
The initial state is the only common 1 eigenvector. 
Common -1 eigenvectors are in particular -1 eigenvectors of the Grover's 
diffusion operator and thus are locally orthogonal to any possible target 
state. Pt. c) is a consequence of the two.
We can now consider the operators $O$ and $SC$ in the $\statespace'$ space so 
that the condition 2 is met.

The proposed method works for any network if the applied operators are 
constructed the same way.

\begin{fact}
The number of steps required to find any marked state using procedure described 
in section \ref{sec:model} adopted for any Apollonian network is $O(\sqrt{N})$.
\end{fact}

We showed that searching for any marked node in the Apollonian network with the 
introduced search model requires $O(\sqrt{N})$ steps. The above results suggest 
that regardless to the 
set selection the search complexity is optimal. In the following section 
we will consider various approaches to the search problem with different sets 
of nodes that can be marked.

\subsection{Measurement step}\label{sec:strategy}

Obtaining an effective search algorithm requires translating the final state 
into classical information about the result. In the case of quantum systems it 
is performed by a measurement. The decision about making a measurement needs 
to be made without any knowledge about the searched node. Thus, it is crucial 
that the designed algorithm enables us to obtain high success probability for 
every possible node in the same step of the evolution.
In the case of the original Grover's search it is satisfied due to completely 
symmetric network structure. When using less complex networks we need to 
ensure that regardless of the localization of the marked node
algorithm efficiency is maintained with use of
exactly the same operations.

\begin{figure*}
\includegraphics[]{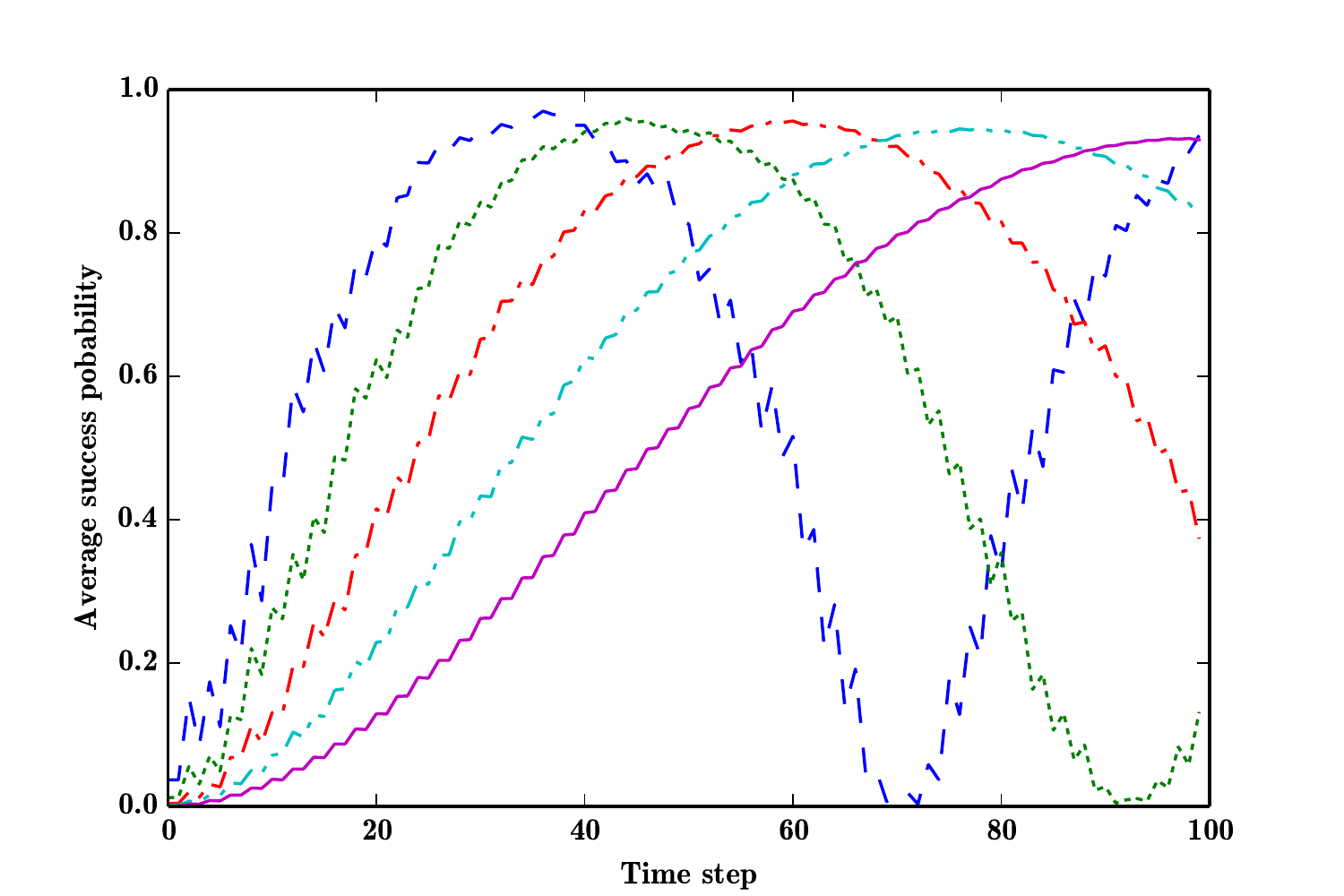}
\caption{Average probability of measuring the marked node grouped by nodes 
generations. Blue, green, red, cyan and magenta lines 
(dashed, dotted, dash-dotted, dash-dot-dotted, solid) correspond 
to nodes created in 4, 5, 6, 7, 8 iteration respectively in the 8th 
generation network.
\label{fig:avg-prob-by-gen}}
\end{figure*}

In this work we aim for analysis of iteratively generated planar networks.
We showed that the 
asymptotic complexity of the search is the same for all of the nodes. 
In particular, we provide an example of the 8th generation Apollonian network 
case. Contrary to the asymptotic results, the numerical results show that 
for nodes from different generations of the network the optimal measurement 
time is different.
This is caused by the fact that the asymptotic complexity considers each of the 
positions separately and neglects the constant factor of the overall complexity 
trend.
The numerical results show how this constant factor differs for various classes 
of nodes.
The probability of measuring a marked node
averaged among the nodes from the same generation is presented in 
Fig.~\ref{fig:avg-prob-by-gen}.
In this case the probability dynamics is very similar for nodes form the same 
generation, but the step of optimal measurement for nodes from different 
generations varies significantly.
Thus, it is not possible to fix the 
measurement time such that 
the algorithm works efficiently when marking arbitrary node in the network is 
allowed.

In this work we differentiate the two cases. The search among all of the nodes 
and search among nodes form the last generation. Both cases exhibits the same 
asymptotic complexity but only the later allows us to perform successful search 
with fixed measurement step.

We utilise the fact that for every node created in the 
last iteration the success probability is maximal at the same time 
step and search with the following algorithm:
\begin{enumerate}
\item Prepare a superposition of all nodes from the last generation.
\item Run quantum walk.
\item Perform a projection onto the subspace of last generation nodes.
\item Measure the position register.
\end{enumerate}
Such approach guarantees that in the moment of measurement the 
probability of successful search will be maximized regardless of the choice of 
the marked node. The projection step can be interpreted as a measurement with 
post-selection. The possibility of measuring the desired subspace is 
approximately one half as the amplitudes localizes approximately equally in the 
marked node and its neighbours. Moreover, if the measurement fails the state of 
the 
systems 
almost surely evolves into desired subspace after application of the shift 
operator. Thus the complexity of the procedure is not affected by the 
projection.

\begin{figure*}
\includegraphics[trim=1.1cm 0cm 0cm 0cm]{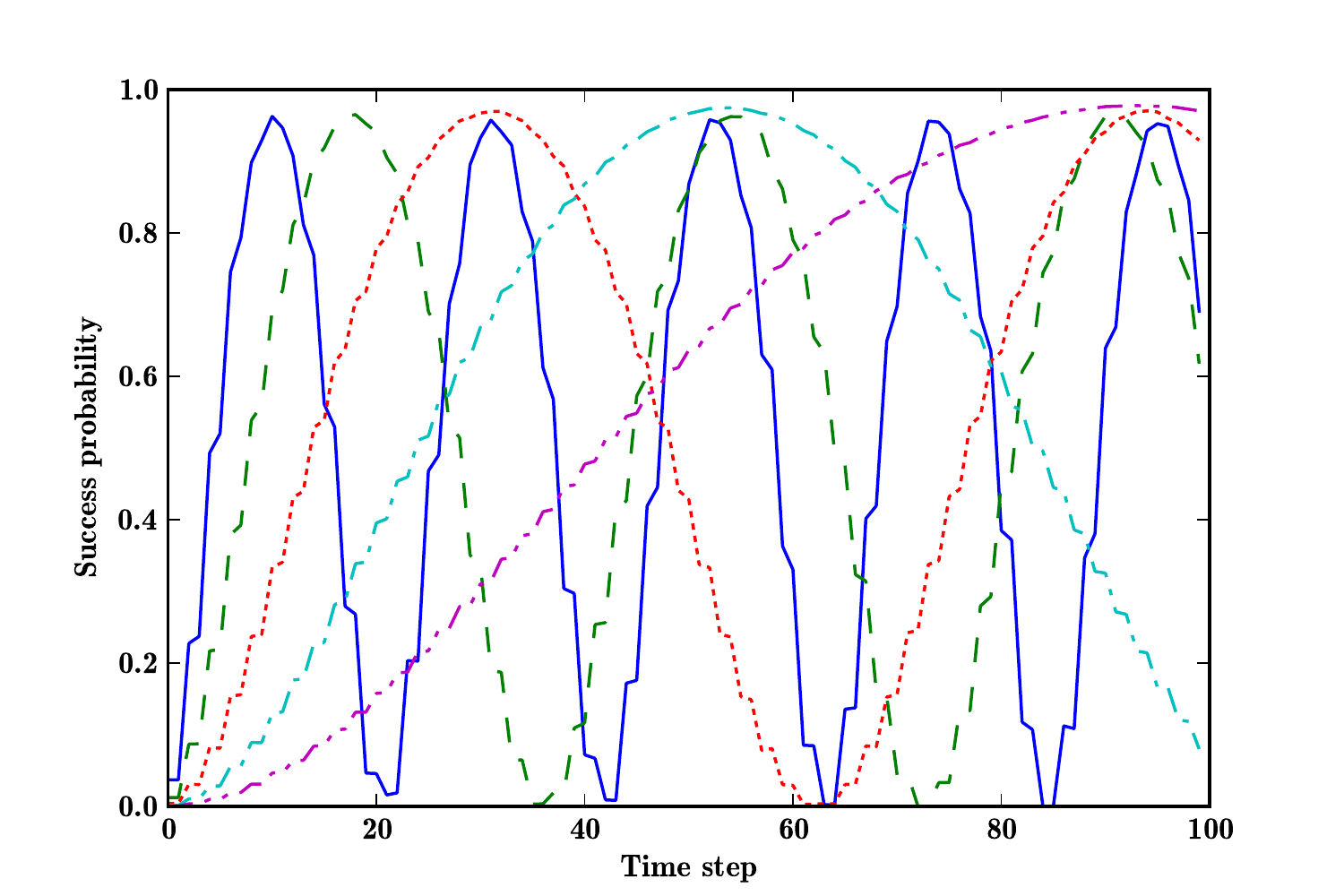}
\caption{Average probability of measuring the searched node (if one of the last 
generation nodes is marked)
in successive steps for network generations: 4 (solid), 5 (dashed),
6 (dotted), 7 (dash-dotted) and 8 (dash-dot-dotted).}
\label{fig:gens_compare}\label{fig:cosinuses}
\end{figure*}

\subsection{Numerical simulation}\label{sec:complexity}

Performing the search tasks as described in section~\ref{sec:strategy} one is 
able to measure the marked node 
with the probabilities presented in Fig. \ref{fig:cosinuses}.
The maximum probability and the corresponding steps are presented in 
Table~\ref{tab:time}.

\begin{table}
\begin{center}
\vspace{0.5cm}
\begin{tabular}{|c|c|c|c|c|c|} \hline
Generation     &    4 &    5 &    6 &    7 &    8  \\ \hline 
$N_{last}$     &   27 &   81 &  243 &  729 & 2187  \\ \hline 
$T_{p}$          &   10 &   18 &   32 &   54 &   92  \\ \hline
$2\sqrt{N_{last}}$ &   10.4 &   18.0 &   31.2 &   54.0 &   93.5  
\\ 
\hline
$\overline{p}(T_p)$       & 0.963 & 0.965 & 0.970 & 0.974 & 0.977  \\ \hline 
\end{tabular}
\end{center}
\caption{The average probability $\overline{p}(T_{p})$ of 
measuring the searched 
node (last generation case) and the
corresponding step $T_{p}$ for Apollonian networks generations 4, 5, 6, 7 
and 8 with number of nodes in the last generation equal to $N_{last}$.}
\label{tab:time}
\end{table}

Let's note that the number of nodes increases exponentially with generation 
number
\begin{equation}
N(K) \approx 3^K.
\end{equation}
As presented in Table \ref{tab:time} the time of maximum success 
probability increases three times every two generations.
For this reason one can assume that the optimal measurement step $T_0$
can be expressed as
\begin{equation}
T_0(K) = \alpha 3^{\frac{K}{2}} = \alpha (\sqrt{3})^K,
\end{equation}
where $\alpha$ is a scaling constant.
With this assumption and using Eq. (\ref{eq:number-of-nodes}) one obtains
\begin{equation}
T_0(N) \approx \alpha \sqrt{N}.
\end{equation}
This allows us to estimate the complexity of the algorithm as
\begin{equation}
E(T_0)= T_0(N)/p(T_0) \approx \alpha \sqrt{N} / p = 
O(\sqrt{N}),
\end{equation}
which induces a quadratic speed-up with respect to the classical algorithm.
Thus, we obtain the optimal speed-up rate for a quantum search algorithm with 
use of an Apollonian network in the case of the restricted search.

\section{Concluding remarks}\label{sec:conclusion}

This work is motivated by the search for simple quantum system structures that 
are sufficient for efficient computation.
We described a quantum walk search procedure on general graph. 
For the purpose of determining network computational potential we applied the 
existing 
asymptotic complexity evaluation methods.
We showed that the asymptotic complexity for the given model is optimal -- 
$O(\sqrt{N})$.
The analytical results show that asymptotic complexity is the same for any 
search scheme.
We focus on showing that the asymptotic complexity is not enough for 
determining the potential of the network.

We perform a numerical study in the case of a quantum walk on Apollonian 
network. We 
considered two cases of the
search. We have developed an algorithm that can be used to search for an
arbitrary node and a node from the last generation of an Apollonian 
network aiming at searching with fixed measurement step. The approach
is not possible to adopt for whole network search. We show numerical evidence
that in the restricted search approach it is possible to adopt Grover's
search algorithm with a measurement time independent from the localization of
the marked node while maintaining the efficiency. Numerical results performed
for generations 4--8, that consist of 43--3283 nodes, allowed us suggesting the
estimate of the algorithm complexity which is consistent with analytical 
predictions. Provided algorithm exhibits quadratic
speed-up in relation to the classical counterpart.

The presented approach exhibits not only quadratic speed-up with respect to the
classical algorithms, but also allows us to obtain quadratic reduction of the 
complexity of the network compared to the basic quantum counterpart by using a 
network with $O(N)$ number of connections instead of $O(N^2)$ in the complete  
network case.

\section{Acknowledgements}
This work is supported by the Polish Ministry of Science and Higher Education
within "Diamond Grant" Programme under the project number 0064/DIA/2013/42 and
by the Polish National Science Centre under the project number
2011/03/D/ST6/00413.

\bibliography{apollonian_grover,../../../../sadowski/citations/walks_search}
\bibliographystyle{apsrev}

\end{document}